\documentclass[11pt]{article}
\usepackage{amsmath,amssymb,color,graphics,epsfig}

\usepackage{amsfonts}

\newcommand{\be}{\begin{equation}}
\newcommand{\ee}{\end{equation}}
\newcommand{\bea}{\setlength\arraycolsep{2pt} \begin{eqnarray}}
\newcommand{\eea}{\end{eqnarray}}
\newcommand{\nn}{\nonumber}

\def\ft#1#2{{\textstyle{\frac{\scriptstyle #1}{\scriptstyle #2} } }}
\def\fft#1#2{{\frac{#1}{#2}}}

\def\0{{\sst{(0)}}}
\def\1{{\sst{(1)}}}
\def\2{{\sst{(2)}}}
\def\3{{\sst{(3)}}}
\def\4{{\sst{(4)}}}
\def\5{{\sst{(5)}}}
\def\6{{\sst{(6)}}}
\def\7{{\sst{(7)}}}
\def\8{{\sst{(8)}}}
\def\sst#1{{\scriptscriptstyle #1}}

\begin{document}

\begin{flushright}
\end{flushright}

\vspace{25pt}
\begin{center}
{\large {\bf Charged Black Holes with Scalar Hair}}

\vspace{10pt}
Zhong-Ying Fan and H. L\"u

\vspace{10pt}

{\it Center for Advanced Quantum Studies\\ }
{\it Department of Physics, Beijing Normal University, Beijing 100875, China}

\vspace{40pt}

\underline{ABSTRACT}
\end{center}

We consider a class of Einstein-Maxwell-Dilaton theories, in which the dilaton coupling to the Maxwell field is not the usual single exponential function, but one with a stationary point.  The theories admit two charged black holes: one is the Reissner-Nordstr\o m (RN) black hole and the other has a varying dilaton. For a given charge, the new black hole in the extremal limit has the same AdS$_2\times$Sphere near-horizon geometry as the RN black hole, but it carries larger mass.  We then introduce some scalar potentials and obtain exact charged AdS black holes.  We also generalize the results to black $p$-branes with scalar hair.

\vfill {\footnotesize Emails: zhyingfan@gmail.com \ \ \ mrhonglu@gmail.com}

\thispagestyle{empty}

\pagebreak

\tableofcontents
\addtocontents{toc}{\protect\setcounter{tocdepth}{2}}




\section{Introduction}

The low-energy effective theories of superstrings, namely supergravities, provide a variety of fundamental matter fields that we can study their interactions with gravity.  These include dilatonic scalars and various form fields.  Under gravity, these fields can give rise to charged black holes or $p$-branes, which are part of fundamental constituents in the perturbative or non-perturbative string spectrum. (See, e.g. \cite{Gibbons:1987ps}.)

   Typically in supergravities, the dilaton $\phi$ couples to an $n$-form field strength $F_n=dA_{n-1}$ through an exponential function.  The relevant Lagrangian is given by
\be
e^{-1} {\cal L} = R - \ft12 (\partial\phi)^2 - \fft1{2\,n!} e^{a\phi} F_n^2\,.
\label{explag}
\ee
where $e=\sqrt{-\det(g_{\mu\nu})}$ and $a$ is the dilaton coupling constant.  The exponential coupling function is determined by the tree-level interaction of the string. The higher-order corrections to supergravity can be extremely complicated, involving both the worldsheet $\alpha'$ and the string coupling constant $g_s=\langle e^{\phi}\rangle$. The general structure was discussed in \cite{Duff:1990hb}. A concrete example is provided by ${\cal N}=1$, $D=6$ supergravity.  In the canonical metric, the dilaton dependence of the kinetic energy of the Yang-Mills gauge field is \cite{Sagnotti:1992qw}
\be
e^{-1} {\cal L}_{\rm gauge} = \sum_\alpha (v_\alpha e^{-\fft1{\sqrt2}\phi} +
\tilde v_{\alpha} e^{\fft1{\sqrt2}\phi}) {\rm tr} F_{\alpha \mu\nu} F_{\alpha}{}^{\mu\nu}\,,\label{d6lag}
\ee
where $(v_\alpha,\tilde v_\alpha)$ are constants, with the former being essentially the Kac-Moody level \cite{Erler:1993zy} and the latter determined by the supersymmetry.  The string one-loop correction $\tilde v_\alpha$-terms play a crucial role \cite{Duff:1996cf} in some heterotic phase transitions \cite{Duff:1996rs,Aldazabal:1996fm,Morrison:1996na}.

     Inspired by (\ref{d6lag}), we consider a general class of Lagrangian
\be
e^{-1} {\cal L} = R - \ft12 (\partial\phi)^2 - \fft1{2\,n!} Z(\phi) F_n^2\,.\label{genlag}
\ee
We require that the function $Z(\phi)$ become an exponential function in either the $\phi\rightarrow +\infty$ or $\phi\rightarrow -\infty$ limit. Thus we can expand $Z$ as ``$e^{a\phi} + \cdots $'' that mimics the string loop expansion. Furthermore, we assume that $Z(\phi)$ has an stationary point, namely $dZ/d\phi=0$ for certain $\phi_0$.
There are two motivations for us to consider such a Lagrangian.  The first is that the black $p$-branes in (\ref{explag}) suffer from curvature singularity on the horizon in the extremal limit, which is typically BPS or  supersymmetric.  In this case, the supergravity solution cannot be trusted to yield some non-perturbative information of strings. By contrast, non-dilatonic $p$-branes have AdS$\times $sphere as their near-horizon geometries in such a limit.  For some appropriate decoupling limit, the smooth classical supergravity background can yield informations of string theory.  This leads to the famous AdS/CFT correspondence \cite{Maldacena:1997re}.

The dilaton coupling function $Z(\phi)$ for (\ref{d6lag}) has a stationary point.  Thus the dilaton can be decoupled.  This implies that the associated $p$-brane can have smooth horizon even in the extremal limit.  It is natural to assume that the singularity associated with the extremal limit of the $p$-branes in (\ref{explag}) is an artifact of the low-energy effect action of the lowest order.  Loop correction is expected to smooth out the singularity and extremal $p$-branes in strings may also have non-vanishing entropy \cite{Lu:1996bd}.  The general loop-corrected actions of strings are very complicated to construct and few examples are known.  The Lagrangian (\ref{genlag}) can be viewed as a toy model for such string loop effects.

The second motivation is to do with uniqueness of black holes.  The standard no-hair theorem states that a black hole is completely specified by the mass, charge and angular momenta.  Recently, many examples of scalar hairy black holes have been constructed [11-26], and their general first law of thermodynamics were derived \cite{Liu:2013gja,Lu:2014maa}. In supergravities, charged black holes also generally involve scalar fields and yet the uniqueness theorem appears to continue to hold for the large number of such solutions constructed in literature. It is of interest to construct a simple counter example.  It is clear that if $Z(\phi)$ in (\ref{genlag}) has a stationary point as in (\ref{d6lag}), then the usual Reissner-Nordstr\"om (RN) black hole with the dilaton decoupled must be a solution.  If we can construct a further different black hole with the same mass and charge, but non-vanishing dilaton, the uniqueness theorem is then broken. The Lagrangian (\ref{genlag}) can thus stand alone as an example for studying non-uniqueness of charged black holes.  Numerically it can be established that charged black holes for a generic $Z(\phi)$ may indeed exist. In this paper, we find non-trivial exact solutions when $Z$ is given by
\be
Z^{-1} = \cos^2\omega\, e^{a_1 \phi} + \sin^2\omega\, e^{a_2\phi}\,,\qquad
\hbox{with}\qquad a_1 a_2 = -\fft{2(n-1)(D-n-1)}{D-2}\,.\label{zcond}
\ee
Here $\omega$ is some coupling constant.  For $\omega=0$ or $\fft12\pi$, the function $Z$ becomes an exponential function of $\phi$.  In general $Z$ has a stationary point for which $dZ/d\phi$ vanishes.  This dilaton coupling to the field strength resembles those in the STU-model truncation \cite{Lu:2014fpa} of $D=4$ $\omega$-deformed gauged supergravity \cite{Dall'Agata:2012bb,deWit:2013ija}.

The paper is organized as follows.  In section 2, we present new charged black holes in four dimensions.  We focus on the examples that are associated with supergravities.  We discuss their global properties and obtain the first law of thermodynamics.  We compare the new black holes with the RN black hole.  We also obtain the multi-center extremal solutions.  We then introduce some scalar potentials and obtain charged AdS black holes.  In section 3, give a construction of the new black holes for general dimensions.  In section 4, we give an alternative construction that allows us to construct $p$-branes with scalar hair in general dimensions.  In section 5, we construct further new charged AdS planar black holes.  We conclude the paper in section 6.

\section{Charged black holes with scalar hair in $D=4$}

In this section, we consider Einstein-Maxwell-Dilaton theory in four dimensions, corresponding to $D=4$ and $n=2$.  The condition for $(a_1,a_2)$ in (\ref{zcond}) becomes $a_1a_2=-1$.  In supergravities with an exponential dilaton coupling, $a_1$ may take values $\pm(\sqrt{3},1,1/\sqrt3,0)$.  We thus first consider the two special cases, namely $(1,-1)$ and $(\sqrt3,-1/\sqrt3)$, in greater detail and summarize the general results later in this section.

\subsection{Case 1: $(a_1,a_2)=(1,-1)$}

We start the discussion by considering an example of charged black holes with scalar hair in four dimensions.  The theory consists of the metric, a dilaton $\phi$ and a Maxwell field $A$.  The Lagrangian is given by
\be
e^{-1} {\cal L} = R - \ft12 (\partial\phi)^2 - \ft14 Z\, F^2\,,\qquad
Z^{-1}=\cos^2\omega\, e^{\phi} + \sin^2\omega\, e^{-\phi}\,.\label{lagd41}
\ee
where $F=dA$.  Naively, the parameter $\omega$ has a range of $\ft12 \pi$; however, owing to the symmetry of $\omega\rightarrow \omega+\ft14\pi$ and $\phi\rightarrow -\phi$, $\omega$ has a range of $\ft14\pi$ instead.  For latter purpose we choose the range of $\omega$ lies in $[\ft14\pi,\ft12\pi]$.  When $\omega=\ft12\pi$, $Z^{-1}$ then involves only one exponential term and the Lagrangian is invariant under the constant shift of the dilaton provided that it is compensated by an appropriate scaling of the gauge potential $A$.  For generic fixed $\omega$, however, such a freedom is deprived.  The $\omega=\ft12\pi$ Lagrangian can be embedded in $D=4$, ${\cal N}=4$ supergravity.

To construct black hole solutions to (\ref{lagd41}), we first note that $Z(\phi)$ has a stationary point for $\omega\ne \ft12\pi$, namely
\be
\fft{dZ}{d\phi}\Big|_{\phi=\phi_0}=0\,,\qquad \hbox{for}\qquad e^{\phi_0}=\tan\omega\,.\label{phi0def1}
\ee
Thus the dilaton can be decoupled, giving rise to the Einstein-Maxwell theory
\be
e^{-1}{\cal L}= R - \fft{1}{4\sin(2\omega)} F^2\,.
\ee
The Lagrangian admits the well-known RN black hole
\bea
ds_4^2 &=& -f dt^2 + \fft{dr^2}{f} + r^2 d\Omega_2^2\,,\qquad
A = \fft{4 \sin(2\omega)\, Q_e}{r} dt\,,\cr
f&=& 1 - \fft{2M}r + \fft{4\sin(2\omega)\, Q_e^2}{r^2}\,.
\eea
where the two integration constants $M$ and $Q_e$ are the mass and electric charge respectively.  For the solution to be a black hole, the mass and charge must satisfy the inequality
\be
M\ge M_{\rm RN}^{\rm ext}\equiv 2\sqrt{\sin(2\omega)}\, Q_e\,.
\ee
The solution becomes extremal when the above inequality is saturated and the black hole has zero temperature with the AdS$_2\times S^2$ near-horizon geometry.

In addition to the RN black hole, we find that the theory (\ref{lagd41}) admits a different charged black hole in which the dilaton is not a constant. We shall give the detail construction later.  For now, we simply present the solution
\bea
ds^2&=&-f dt^2 + \fft{dr^2}{f} + r(r+q) d\Omega_2^2\,,\qquad
f = 1 + \fft{Q^2\big(q + (2r+q)\cos(2\omega)\big)}{4qr(r+q)}\,,\cr
e^\phi&=&1 + \fft{q}{r}\,,\qquad A=\fft{Q(r+q \cos^2\omega)}{r(r+q)}dt\,,
\eea
The solution involves two integration constant $q$ and $Q$, which, as we shall see presently, parameterize the mass and electric charge of the solution. We assume that $q>0$. With this choice, the spacetime has no curvature singularity in the $r>0$ region.  For the solution to describe a black hole, the curvature singularity at $r=0$ has to be shielded by an event horizon at certain $r=r_0>0$; the curvature singularity at $r=-q<0$ is irrelevant for our discussion.

In the large $r$ expansion, we have
\be
-g_{tt}=1 + \fft{Q^2\cos(2\omega)}{2q R} + \fft{Q^2}{4R^2} + \cdots\,,
\label{finfinity}
\ee
where $R=\sqrt{r(r+q)}$ is radius of the foliating 2-sphere.  We can read off the ADM mass
\be
M=-\ft14 \cos(2\omega)\, \fft{Q^2}{q}\,.
\ee
We see now the reason that we have chosen $\omega\in [\ft14\pi,\ft12\pi]$.  (Of course, one can perfectly choose $\omega\in [0,\ft14\pi]$, in which case, we can simply let $q$ be negative and the curvature singularity $r=-q>0$ becomes the relevant one to discuss.) The electric charge of the solution is given by
\be
Q_e = \fft{1}{16\pi} \int_{r\rightarrow\infty} Z {*F}=\ft14 Q\,.
\ee
To determine the relation of the mass $M$ and $Q_e$ so that the solution describe a black hole, we note that the function $f$ approaches $+\infty$ at $r=0$ and becomes 1 as $r\rightarrow +\infty$.  It follows from (\ref{finfinity}) that for positive $M$, there must be a minimum for $f$ at some $r=r_{\rm min}>0$.  It is straightforward to
determine that
\be
r_{\rm min}=\fft{q\cos \omega}{\sin\omega-\cos\omega}\,,\qquad f^{\rm min}=f(r_{\rm min}) = \fft{1}{2q^2} \big(Q^2\sin (2\omega) -Q^2 + 2q^2\big)\,.
\ee
Thus the solution describes a black hole provided that $f^{\rm min}\le 0$, i.e.
\be
\sin(2\omega) \le 1 - \fft{2q^2}{Q^2}\,.
\ee
When the inequality is satisfied, $f$ has two real positive roots
\be
r_\pm = \fft{1}{4q}\Big(\pm\sqrt{(Q^2-2q^2)^2 -Q^4\sin^2(2\omega)}-2q^2 - Q^2\cos(2\omega)\Big)\,.
\ee
Using the standard technique, we obtain the temperature and the entropy
\be
T=\fft{2(r_0+q)^2 \cos^2 \omega -2 r_0^2 \sin^2\omega}{4\pi r_0(r_0+q)
\big(q + (2r_0+q) \cos(2\omega)\big)}\,,\qquad
S=r_0(r_0+q) \pi\,.
\ee
where $r_0=r_+$. The electric potential difference between the horizon and the asymptotic infinity is
\be
\Phi_e=\fft{Q(r_0+q \cos^2\omega)}{r_0(r_0+q)}\,.
\ee
It is then straightforward to verify that the first law of black hole thermodynamics
\be
dM=TdS + \Phi_e dQ_e\label{firstlaw}
\ee
is satisfied.  These quantities satisfy also the Smarr relation
\be
M= 2 TS + \Phi_e Q_e\,.\label{smarr}
\ee
Thus we see that for given mass $M$ and electric charge $Q_e$, the theory (\ref{lagd41}) admits two black holes, one is the RN black hole with constant $\phi$, whilst the other is the scalar hairy one involving a varying $\phi$.

The new scalar hairy solution also has an extremal limit in which $T$ vanishes, corresponding to
\be
\tan\omega = \fft{r_0+q}{r_0}=e^{\phi}\Big |_{r=r_0}\,.
\ee
Thus we see that $\phi|_{r=r_0}=\phi_0$, where $\phi_0$ was defined in (\ref{phi0def1}). The mass of the extremal black hole with scalar hair is then given by
\be
M_{\rm SH}^{\rm ext}=\sqrt2 (\cos\omega+ \sin\omega)Q_e\,.
\ee
It is easy to verify that for the same electric charge, the mass of the extremal hairy black hole is in general bigger than the extremal RN one, i.e.
\be
M_{\rm SH}^{\rm ext} \ge M_{\rm RN}^{\rm ext}\,.\label{inequal1}
\ee
In particular, this implies that in the extremal limit, both the new and RN black holes with the same electric charge have the same AdS$_2\times S^2$ structure as their near horizon geometry, but the asymptotic behaviors are different; the new solution having a larger mass. The inequality (\ref{inequal1}) is saturated for $\omega=\ft14\pi$ for which the scalar hairy black hole degenerates to become the RN black hole.

   Finally, we note that both the new and the RN black holes have two horizons, inner and outer horizon.  The product of the entropies of the two horizons are quantized in the same way
\be
S_+ S_- = 16\pi^2 Q_e^4 \sin^2(2\omega)\,.
\ee
The universality of this property of two very different solutions of the theory is rather telling that there may be a same conformal field theory that is responsible for the interpretation of the black hole entropy.  (See, e.g.~\cite{CYII,L}.) This entropy product rule also implies that although for given electric charge $Q_e$, the mass of RN and the new scalar hairy black hole is different, but their entropy is the same in the extremal limit.

\subsection{Case 2: $(a_1,a_2)=(\sqrt3,-1/\sqrt3)$}

The Lagrangian for the second example takes the same form as (\ref{lagd41}), but now with $Z$ given by
\be
Z^{-1} = \cos^2\omega\, e^{\sqrt3\,\phi} + \sin^2\omega\, e^{-\fft{1}{\sqrt3}\phi}\,.
\ee
The range of $\omega$ lies in the interval $[0,\fft12\pi]$.  We pick up this example to discuss in some detail because the theory can be embedded in supergravity when $\omega=0$ or $\fft12\pi$. The function $Z$ has a stationary point $\phi=\phi_0$ with  \be
e^{\fft{4}{\sqrt3}\phi_0}=\ft13\tan^2\omega\,.
\ee
Thus the dilaton can be decoupled, giving rise to Einstein Maxwell theory
\be
e^{-1}{\cal L} = R - \fft1{4\Theta^2} F^2\,,\label{einmax}
\ee
where
\be
\Theta=2\cdot 3^{-\fft38}\, (\cos\omega)^{\fft14} (\sin\omega)^{\fft34}\,.
\ee
Thus the theory admits the RN black hole and its mass and charge satisfy
\be
M\ge M_{\rm RN}^{\rm ext} = 2\Theta\, Q_e\,.
\ee
The bound is saturated in the extremal limit.

We find that the charged scalar hairy black hole is given by
\bea
ds^2 &=& - f dt^2 + \fft{dr^2}{f} + r^2(1 + \fft{q}{r})^{\fft32} d\Omega_2^2\,,
\qquad A=\fft{Q(r+q \cos^2\omega)}{r(r+q)}\,,\cr
e^{\fft{2}{\sqrt3}\phi} &=& 1 + \fft{q}{r}\,,\qquad
f=\fft{1}{\sqrt{r(r+q)^3}}\Big(r(r+q) + \fft{Q^2}{3q} \big(
3(r+q) \cos^2\omega - r \sin^2\omega\big)\Big)\,.
\eea
Following the same technique in the earlier discussion, we find that the mass and the electric charge of the black hole is given by
\be
M=\fft{1}{12q}\big( 3q^2 + 2Q^2 - 8 Q^2 \cos^2\omega\big)\,,\qquad
Q_e=\ft14 Q\,.
\ee
It is easy to establish that when the parameters satisfy
\be
(2Q\cos\omega - \ft32 q)^2 \ge \ft34 q^2 - Q^2\,,\label{bhcond2}
\ee
the function $f(r)$ has two positive roots $r_\pm>0$.  The horizon is located at $r_0\equiv r_+\ge r_-$.  The temperature, the entropy, and electric potential are
\bea
T &=& \fft{3(r_0+q)^2\cos^2\omega - r_0^2\sin^2\omega}{4\pi\sqrt{r_0(r_0+q)^3}
\big(3(r_0+q) \cos\omega^2 - r_0\sin^2\omega\big)}\,,\cr
S &=& \pi \sqrt{r_0(r_0+q)^3}\,,\qquad \Phi_e=\fft{Q(r_0+q \cos^2\omega)}{r_0(r_0+q)}\,.
\eea
We verify that both the first law (\ref{firstlaw}) and the Smarr relation (\ref{smarr}) are satisfied.  In the extremal limit, where the temperature vanishes, the mass is proportional to the electric charge
\be
M_{\rm SH}^{\rm ext} = (\cos\omega + \sqrt3 \sin\omega)Q_e\,.
\ee
The condition (\ref{bhcond2}) for the existence of event horizon is equivalent to $M\ge M_{\rm SH}^{\rm ext}$.  In general $M_{\rm SH}^{\rm ext}\ge M_{\rm RN}^{\rm ext}$ with the equality is achieved at $\omega=\ft13\pi$, in which case, the hairy black hole degenerates to the RN black hole.

\subsection{The general case}

We now consider the general case with $a_1a_2=-1$.  We reparameterize the constants
as
\be
a_1=\sqrt{\fft{1-\mu}{1+\mu}}\,,\qquad a_2=-\sqrt{\fft{1+\mu}{1-\mu}}\,.
\ee
The dilaton coupling function $z$ is thus given by
\be
Z^{-1}= \cos^2\omega\, e^{\sqrt{\fft{1-\mu}{1+\mu}}\,\phi} +
\sin^2\omega\, e^{-\sqrt{\fft{1+\mu}{1-\mu}}\,\phi}\,,\label{d4Zgen}
\ee
It has a stationary point, given by
\be
e^{\fft{1}{\sqrt{1-\mu^2}}\phi_0} = \sqrt{\ft{1+\mu}{1-\mu}}\,\tan\omega\,.
\ee
The reduced Einstein-Maxwell theory (\ref{einmax}) has
\be
\Theta=\fft{\sin(2\omega)}{\sqrt{1-\mu^2}} \Big(\sqrt{\ft{1+\mu}{1-\mu}}\,\cot\omega\Big)^\mu\,.
\ee
We find that the scalar hairy black hole is given by
\bea
ds^2 &=& -f dt^2 + \fft{dr^2}{f} + r^{1+\mu}(r + q)^{1-\mu} d\Omega_2^2\,,\qquad A=\fft{Q(r+q \cos^2\omega)}{r(r+q)}\,,\cr
e^{\fft{1}{\sqrt{1-\mu^2}}\phi} &=& 1 + \fft{q}{r}\,,\qquad
f=(1 + \fft{q}{r})^\mu\Big(1 + \fft{\cos^2\omega\, Q^2}{2(1+\mu)q r} -
\fft{\sin^2\omega\, Q^2}{2(1-\mu)q(r+q)}\Big)\,.
\eea
The mass and electric charge are given by
\be
M=\fft{Q^2\sin^2\omega}{4(1-\mu)q} -\fft{Q^2\cos^2\omega}{4(1+\mu)q} -\ft12\mu q\,,\qquad
Q_e=\ft14 Q\,.
\ee
For appropriate parameter regions, the event horizon exists and the first law (\ref{firstlaw}) and smarr relation (\ref{smarr}) can be shown to be satisfied.  For a given electric charge $Q_e$, the black hole mass must satisfy the inequality
\be
M\ge M^{\rm ext}_{\rm SH}\equiv \sqrt2(\sqrt{1 + \mu} \cos\omega +
\sqrt{1-\mu} \sin\omega)Q_e\,.
\ee
When the equality is achieved, the black hole becomes extremal with zero temperature and the corresponding near horizon geometry is AdS$_2\times S^2$.  For the extremal solution, the dilaton runs from $\phi=\phi_0$ on the horizon to $\phi=0$ to asymptotic flat infinity.  For a given electric charge $Q_e$, the mass of extremal solution of scalar hairy black hole is in general bigger than that of the RN back hole, namely
\be
M^{\rm ext}_{\rm SH} \ge M^{\rm ext}_{\rm RN}\,.
\ee
The equality occurs when $\tan\omega = \sqrt{1-\mu}/\sqrt{1+\mu}$ in which case, the scalar hairy black hole degenerates to the RN black hole.

The general black holes have two horizons, one is outer and the the other is inner. The product of the two horizon entropies is quantized:
\be
S_+ S_- =64\pi^2\big(\fft{\cos^2\omega}{1+\mu}\big)^{1+\mu}
\big(\fft{\sin^2\omega}{1-\mu}\big)^{1-\mu}\,Q_e^4 \,.
\ee

\subsection{The extremal limit and multi-centered black holes}

As in the case of RN black holes, in the extremal limit, the electric force and gravity balance out and one can have multi-centered solutions.  We find that the multi-centered solutions for the general $Z$ (\ref{d4Zgen}) are given by
\bea
ds_4^2 &=& H_1^{-1+\mu}H_2^{-1+\mu} dt^2 + H_1^{1+\mu}H_2^{1-\mu} dx^i dx^i\,,\cr
A &=& \big(\fft{\sqrt{2(1+\mu)}\,\cos\omega}{H_1} + \fft{\sqrt{2(1-\mu)}\,\sin\omega}{H_2}\big) dt\,,\qquad \phi=\sqrt{1-\mu^2} \log \fft{H_2}{H_1}\,,\cr
H_1 &=& 1+ \fft{\cos\omega}{\sqrt{2(1-\mu)}}\, H_0\,,\qquad
H_2=1 + \fft{\sin\omega}{\sqrt{2(1+\mu)}}\, H_0\,,
\eea
where $H_0$ is given by
\be
H_0=\sum_i \fft{Q_i}{|\vec x-\vec x_i|}\,.
\ee
The solution reduces to AdS$_2\times S^2$ near the location of each charge, namely
$\vec x\rightarrow \vec x_i$.

\subsection{Adding a scalar potential}

The charged black hole solutions we considered are all asymptotic to the flat Minkowski spacetime.  We would like to add a scalar potential to the Lagrangian and construct charged black holes that are asymptotic to AdS.

For $\omega=0$ or $\omega=\fft12\pi$, charged AdS black holes were previously constructed. The scalar potential is given by
\be
V(\phi)=-(g^2-\alpha) V_0(\phi) - \alpha V_0(-\phi)\,,
\ee
with
\be
V_0(\phi) = (1+\mu)(1+2\mu) e^{-\sqrt{\fft{1-\mu}{1+\mu}}\,\phi} +
(1-\mu)(1-2\mu)  e^{\sqrt{\fft{1+\mu}{1-\mu}}\,\phi} + 4(1-\mu^2) e^{\fft{\mu}{\sqrt{1-\mu^2}}\,\phi}\,.
\ee
We find that the Lagrangian admits charged black hole solution for general $\omega$.
The solution is given by
\bea
ds^2 &=& -f dt^2 + \fft{dr^2}{f} + r^{1+\mu}(r + q)^{1-\mu} d\Omega_{2,k}^2\,,\cr
A&=&\fft{Q(r+q \cos^2\omega)}{r(r+q)}\,,\qquad e^{\fft{1}{\sqrt{1-\mu^2}}\phi} = 1 + \fft{q}{r}\,,
\eea
with
\bea
f &=& (g^2-\alpha) \big(1 + \fft{q}{r}\big)^\mu \big(r(r+q) - \mu q (2r+q) + 2\mu^2 q^2\big) + \alpha r^2 \big(1+\fft{q}{r}\big)^{1-\mu}\cr
&& + (1 + \fft{q}{r})^\mu\Big(k + \fft{\cos^2\omega\, Q^2}{2(1+\mu)q r} -
\fft{\sin^2\omega\, Q^2}{2(1-\mu)q(r+q)}\Big)\,.
\eea
Here $k$ takes the values 1,0 or $-1$ corresponding to the unit round sphere, torus or hyperbolic space for $d\Omega_{2,k}^2$.

     It is worth commenting that although both $V$ and $Z$ have stationary points, they do not coincide.  Thus the dilaton cannot be decoupled and the RN-AdS black hole is not a solution of the theory.

\section{The construction in general dimensions}

In the previous section, we presented a class of charged black holes with scalar hair in four dimensions. In this section, we show the method of construction.  We give the construction in general dimensions, but continue to focus on black holes. We shall present an alternative construction in the next section where we obtain a class of black $p$-branes with scalar hair.

\subsection{Asymptotically-flat black holes}

We now consider Einstein gravity coupled to a Maxwell field with a generic dilaton coupling in general $D$ dimensions
\be
e^{-1}{\cal L}_D = R - \ft12(\partial\phi)^2 - \ft14 Z(\phi)\, F^2\,,
\ee
The equations of motion are
\bea
\Box \phi &=& \ft14 \ft{dZ}{d\phi}\, F^2\,,\qquad
\nabla_\mu (Z F^{\mu\nu}) =0\,,\cr
E_{\mu\nu} &\equiv& R_{\mu\nu}-\ft12\partial_\mu\phi\partial_\nu\phi - \ft12 Z\big(F_{\mu\nu}^2 - \ft1{2(D-2)} F^2 g_{\mu\nu}\big)=0\,,
\eea
We consider the ansatz
\be
ds^2 = - H^{-1} f dt^2 + H^{\fft1{D-3}} \big(\fft{dr^2}{f} + r^2 d\Omega_{D-2,k}^2\big)\,,\qquad A=\psi\,dt\,.\label{gendmetric}
\ee
The functions $H,f,\psi$ and the dilaton $\phi$ are all functions of $r$ only.  The metric $d\Omega_{D-2,k}^2$ with $k=1,0,-1$ denotes that of the maximally symmetric space with $R_{ij}=(D-3)k\delta_{ij}$.  In other words, it is for round sphere, torus, or hyperbolic $(D-2)$-space respectively.  For asymptotic flat black holes, we must have $k=1$, but we keep $k$ generic for discussing local solutions.

The equation of motion of $A$ implies that
\be
\psi'=\fft{Q}{ZH\, r^{D-2}}\,,
\ee
where $Q$ is an integration constant related to the electric charge
\be
Q_e = \fft{1}{16\pi} \int_{r\rightarrow \infty} Z\, {*F}=\fft{Q\Sigma_{D-2}}{16\pi}\,,
\ee
where $\Sigma_{D-2}$ is the volume of the space associated with $d\Omega_{D-2,k}^2$.
This is the most general ansatz that respects the isometry and it was first considered in \cite{Feng:2013tza}.  A nice property of this ansatz is that the combination of the Einstein equations $E_t{}^t - (D-3) E_{i}{}^i$ yields
\be
r^2 f'' + (3D-8) r f' + 2(D-3)^2 (f-k)=0\,.
\ee
Thus we have
\be
f=k + \fft{c_1}{r^{D-3}} + \fft{c_2}{r^{2(D-3)}}\,,\label{fres1}
\ee
where $c_1, c_2$ are two integration constants.  The combination $E_{t}{}^t-E_r{}^r$ gives
\be
\fft{2(D-3)}{D-2} \phi'^2 + \fft{2H''}{H} -\fft{H'^2}{H^2} + \fft{2(D-2)H'}{rH}=0\,.
\ee
A further ansatz is thus necessary for obtaining exact solutions.  We follow \cite{Feng:2013tza} and assume the dilaton is given by
\be
\phi=\sqrt{\ft{D-2}{2(D-3)}} \sqrt{1-\mu^2} \log \fft{H_1}{H_2}\,,\qquad H_i=1 + \fft{q_i}{r^{D-3}}\,,\qquad i=1,2\,.\label{scalarans1}
\ee
Then the function $H$ can be solved exactly, namely \cite{Feng:2013tza}
\be
H=H_1^{1+\mu} H_2^{1+\mu}\big(c H_1^{-\mu} + (1-c) H_2^{-\mu}\big)^2\,,\label{genH}
\ee
where $c$ is an integration constant.  (Another integration constant is overall scale of $H$ which we fix so that $H$ becomes unity as $r\rightarrow \infty$.)  We can then read off the function $Z$ as a function $r$ with the remainder of the Einstein equations.  Using (\ref{scalarans1}) we can rewrite $Z$ in terms of $\phi$.  We would like to insist that $Z(\phi)$ is independent of the parameters $q$ and $Q$, this implies that we must have $c=0$ or $c=1$ for generic $k$.  (New possibility arises for $k=0$, which we shall discussion in section 5.)  The choice of $c=0$ and $c=1$ is simply an interchange $q_1$ and $q_2$ and hence we shall choose without loss of generality $c=0$, and hence
\be
H=H_1^{1+\mu} H_2^{1-\mu}\,.
\ee
The function $Z(\phi)$ is then given by
\bea
Z^{-1} &=&\cos^2\omega\, e^{a_1 \phi} + \sin^2\omega\, e^{a_2\phi}\,,\cr
a_1 &=&\sqrt{\ft{2(D-3)}{D-2}} \sqrt{\ft{1+\mu}{1-\mu}}\,,\qquad
a_2=-\sqrt{\ft{2(D-3)}{D-2}} \sqrt{\ft{1-\mu}{1+\mu}}\,.\label{zfunctionbh}
\eea
Note that we have chosen the constants $c_1$ and $c_2$ in (\ref{fres1}) such that the parameters $q_i$ and $Q$ does not appear in $Z(\phi)$ explicitly.  The metric function $f$ is now given by
\be
f=k\, H_1 H_2 + \fft{Q^2}{(D-2)(D-3)(q_1-q_2) r^{D-3}}\Big(
\fft{\cos^2\omega}{1-\mu} H_1 - \fft{\sin^2\omega}{1+\mu} H_2\Big)\,.
\ee
Thus the theory and the corresponding black hole solution are both fully determined.

The function $Z$ has a stationary point and hence the dilaton can be decoupled, giving rise to Einstein-Maxwell theory (\ref{einmax}) with
\be
\Theta=2 \Big(\fft{\cos^2\omega}{1-\mu}\Big)^{\fft12(1-\mu)}
\Big(\fft{\sin^2\omega}{1+\mu}\Big)^{\fft12(1+\mu)}\,.
\ee
This quantity turns out to be dimensionally independent.  At large $r$, we find that
\be
g_{tt} =-k + \Big(k \mu q -\fft{Q^2(\mu + \cos(2\omega))}{(D-2)(D-3)(1-\mu^2)q}\Big) \fft{1}{R^{D-3}}-\fft{Q^2}{2(D-2)(D-3)R^{2(D-3)}}+\cdots\,,
\ee
where $q=q_1-q_2$ and $R=r H^{\fft{1}{2(D-3)}}$.  Thus the ADM mass is given by
\be
M=\fft{\Sigma}{16\pi} \Big((D-2)k \mu q -\fft{Q^2(\mu + \cos(2\omega))}{(D-3)(1-\mu^2)q}\Big)\,.
\ee
Here it is understood that $k=1$.  The Maxwell field is given by
\be
A=-\fft{Q}{D-3}\Big(\fft{\cos^2\omega}{r^{D-3} + q_2} + \fft{\sin^2\omega}{r^{D-3} + q_1}\Big)dt\,.
\ee
It follows that the electric charges and potential are given by
\be
Q_e= \fft{Q}{16\pi} \Sigma\,,\qquad
\Phi_e=\fft{Q}{D-3}\Big(\fft{\cos^2\omega}{r_0^{D-3} + q_2} + \fft{\sin^2\omega}{r_0^{D-3} + q_1}\Big)\,,
\ee
where $r_0$ is the location of the horizon.  With the standard technique for calculating the temperature and entropy $T,S$, we can easily verify the first law of thermodynamics and the Smarr relation, given by
\be
dM=TdS + \Phi_e dQ_e\,,\qquad M=\fft{D-2}{D-3} TS + \Phi_e Q_e\,.\label{usualsmarr}
\ee
The general solution also has two horizons with the product of the two corresponding entropies being quantized, namely
\be
S_+ S_- = \ft1{16}\Sigma^2 \Big(
\fft{Q^2}{(D-2)(D-3)} \big(\fft{\cos^2\omega}{1-\mu}\big)^{1-\mu}
\big(\fft{\sin^2\omega}{1+\mu}\big)^{1+\mu}\Big)^{\fft{D-2}{D-3}}\,.
\ee

\subsection{Asymptotic AdS solutions}

We now consider adding a scalar potential $V$ to the Lagrangian.  For the metric ansatz (\ref{gendmetric}), the results were already obtained in \cite{Feng:2013tza}.  The scalar potential turns out to be
\bea
V &=& - \ft12 (D-2) g^2 e^{\fft{\mu-1}{\nu}\Phi}\Big[
(\mu-1)\big((D-2)\mu-1\big) e^{\fft{2}{\nu}\Phi} -2(D-2)(\mu^2-1) e^{\fft{1}{\nu}\Phi}\cr
&&\qquad\qquad\qquad\qquad\quad + (\mu+1)\big((D-2)\mu+1\big)\Big]\cr
&&-\ft{(D-3)^2}{2(3D-7)}(\mu+1) \alpha\, e^{-\fft{1}{\nu} (4+\fft{\mu+1}{D-3})\Phi} (e^{\fft{1}{\nu}\Phi} -1)^{3+\fft{2}{D-3}}\cr
&&\times\Big[(3D-7) e^{\fft{1}{\nu}\Phi}\,{}_2F_1[2,1+\ft{(D-2)(\mu+1)}{D-3};
3+\ft{2}{D-2};1 - e^{\fft1{\nu}\Phi}] +\cr
&&\quad - \big((3D-7) + (D-2)(\mu-1)\big)\,
{}_2F_1[3,2+\ft{(D-2)(\mu+1)}{D-3};4+\ft{2}{D-2};1 - e^{\fft1{\nu}\Phi}]
\Big]\,,\label{genpot}
\eea
where $\nu=\sqrt{1-\mu^2}$ and
\be
\Phi=\sqrt{\ft{2(D-3)}{D-2}}\,\phi\,.\label{Phidef}
\ee
The solution is then given by
\bea
f &\rightarrow & f + g^2 r^2 \left(H_1^{1+\mu} H_2^{1-\mu}\right)^{\fft{D-2}{D-3}}\cr
&&-\alpha\, r^2 H_2 (H_1-H_2)^{\fft{D-1}{D-3}}\, {}_2F_1[1, \ft{D-2}{D-3}(1+\mu); \ft{2(D-2)}{D-3}; 1 - \ft{H_2}{H_1}]\,.
\label{fsol1}
\eea
Since now the solutions are asymptotic AdS, we can let $k=1,0,-1$ instead of only $k=1$ in the asymptotic flat case.

\section{Alternative construction and scalar hairy black $p$-branes}

In the previous section, we construct a class of charged scalar hairy black holes that are asymptotic to flat spacetimes.  The $Z$ function is given by (\ref{zfunctionbh}) with
\be
a_1 a_2 =-\fft{2(D-3)}{D-2}\,.
\ee
This is precisely a condition that one can construct analytical two-charge black holes for the Lagrangian \cite{Lu:2013eoa}:
\be
e^{-1} {\cal L}=R - \ft12 (\partial\phi)^2 - \fft14 e^{a_1 \phi} (F^1)^2 -
\fft14 e^{a_2 \phi} (F^2)^2\,.
\ee
This suggests that a relation exists between our solution and the two-charge solutions constructed in \cite{Lu:2013eoa}.

To establish an even more general relation, we consider a Lagrangian involving two $\tilde n=D-n$ form field strengths $F^i_{\tilde n} = dA_{\tilde n-1}^i$ in $D$-dimensions
\be
e^{-1} {\cal L} = R - \ft12 (\partial\phi)^2 - \fft{1}{2\ \tilde n!} e^{a_1 \phi}
(F_{\tilde n}^1)^2 - \fft{1}{2\ \tilde n!} e^{a_2 \phi} (F_{\tilde n}^2)^2 \,,
\ee
The dilaton coupling constants $(a_1,a_2)$ satisfy
\be
a_1 a_2 = - \fft{2(n-1)(D-n-1)}{D-2}\,.\label{intersectioncond}
\ee
We can parameterize the $a_1$ and $a_2$ as following
\be
a_1^2 = \fft{4}{N_1} + a_1 a_2\,,\qquad a_2^2 = \fft{4}{N_2} + a_1 a_2\,,
\ee
In this parametrization, we have the following two identities
\be
a_1 N_1 + a_2 N_2 =0\,,\qquad a_1 a_2 (N_1+N_2) = -4\,.
\ee
The Lagrangian admits the $(n-2)$-brane with magnetic charges.  The metric and the dilaton are given by (see e.g.~\cite{Lu:1995cs,Duff:1996hp,Cvetic:1996gq},)
\bea
ds^2 &=& \big(H_1^{N_1} H_2^{N_2}\big)^{-\fft{D-n-1}{D-2}} \big(-f dt^2 + dx^i dx^i
\big)+ \big(H_1^{N_1} H_2^{N_2}\big)^{\fft{n-1}{D-2}} \big(\fft{dr^2}{f} + r^2 d\Omega_{D-n}^2\big)\,,\cr
\phi &=& \ft12 a_1 N_1 \log H_1 + \ft12 a_2 N_2 \log H_2\,,\cr
f &=& 1- \fft{\mu}{r^{D-n-1}}\,,\qquad H_i = 1 + \fft{\mu\sinh\delta_i^2}{r^{D-n-1}}\,,\qquad i=1,2\,.\label{p-brane}
\eea
The gauge fields are given by their field strengths
\be
F_{\tilde n}^i = P_i\, \Omega_{D-n}\,,\qquad P_i=\ft12\,(D-n-1) \mu \sqrt{N_i} \sinh(2\delta_i)\,.
\ee
The solutions have three integration constants $(\mu,\delta_1,\delta_2)$, giving rise to the mass and two magnetic charges.  We now redefine the parameters to be
\be
P_1=\cos^2\omega\, Q\,,\qquad P_2=\sin^2\omega\,Q\,,
\ee
The corresponding solution can be also obtained from the Lagrangian
\be
e^{-1} {\cal L} = R - \ft12 (\partial\phi)^2 - \fft{1}{2\ \tilde n!} (\cos^2\omega\, e^{a_1 \phi}+\sin^2\omega\, e^{a_2\phi})
(F_{\tilde n})^2\,.
\ee
The corresponding field strength is given by
\be
F_{\tilde n}=Q\, \Omega_{D-2}\,.
\ee
We perform electric and magnetic duality on the $\tilde n$-form field strength to become an $n$-form $F_n=dA_{n-1}$.  The corresponding theory becomes
\be
e^{-1} {\cal L} = R - \ft12 (\partial\phi)^2 - \fft{1}{2\ n!} Z(\phi) F_n^2\,,\qquad Z^{-1} =\cos^2\omega\, e^{a_1 \phi}+\sin^2\omega\, e^{a_2\phi}\,.
\ee
The electrically charged black $p$-brane solution is then given by (\ref{p-brane}), but with the field gauge potential
\be
A_{n-1}=a\, dt\wedge dx^1\wedge\cdots \wedge dx^{n-2}\,,\qquad
a'=\fft{Q}{Zr^{D-n-2}} \big(H_1^{N_1} H_2^{N_2}\big)^{\fft12 a_1 a_2}\,.
\ee
The solution now involves two integration constants $(\mu,Q)$, giving rise to the mass and electric charge.  The parameter $\omega$ is now a coupling constant in the theory.  It should be pointed out again that the theory admits also a non-dilatonic $p$-brane carrying the same mass and charge where the dilaton is a constant, as in the case of the black hole examples discussed earlier.

The multi-center $(n-1)$-brane solution in the extremal limit can also be easily obtained, given by
\bea
ds^2 &=& \big(H_1^{N_1} H_2^{N_2}\big)^{-\fft{D-n-1}{D-2}} dx^\mu dx^\nu\eta_{\mu\nu}+ \big(H_1^{N_1} H_2^{N_2}\big)^{\fft{n-1}{D-2}} dy^mdy^m\,,\cr
\phi &=& \ft12 a_1 N_1 \log H_1 + \ft12 a_2 N_2 \log H_2\,,\cr
A_n&=& \big(\fft{\sqrt{N_1}\cos\omega}{H_1} + \fft{\sqrt{N_2}\sin\omega}{H_2}\big)dt\wedge dx^1\cdots\wedge dx^{n-1}\,,\cr
H_1&=&1 + \fft{\cos\omega}{\sqrt{N_1}} H_0\,,\qquad
H_2=1 + \fft{\sin\omega}{\sqrt{N_2}} H_0\,,\label{multi-center}
\eea
where
\be
H_0 =\sum_\alpha \fft{Q_\alpha}{|\vec y-\vec y_\alpha|^{D-n-1}}\,.
\ee
The geometry around each center of the charges is AdS$_n\times S^{D-n}$.

\section{New charged AdS planar black holes}

As was discussed under (\ref{genH}) in section 3, we could construct more possibilities of the $Z$ function when the spatial section is flat, i.e.~$k=0$.  Although for a black hole to be asymptotic flat, we must have $k=1$, the $k=0$ case is allowed for asymptotic AdS solutions. In this section, we consider again the EMD theory with general coupling function $Z$ and scalar potential $V$.

\subsection{Class 1}

The class-1 solutions are specified by the metric ansatz
\be
ds^2 = -f dt^2 + \fft{dr^2}{f} + r^{1+\mu} (r+q)^{1-\mu} dx^i dx^i\,.
\ee
Large classes of neutral scalar hairy AdS planar black holes were obtained in \cite{Acena:2013jya,Fan:2015tua}. The scalar potential is
\bea
V(\phi)&&=\ft 14(D-2)g^2e^{-\lambda \mu\phi}\Big(2D(\mu^2-1)+(2-D+D\mu)(1-\mu)e^{-\lambda\phi}\cr
&&+(2-D-D\mu)(1+\mu)e^{\lambda\phi} \Big)  -\ft 14(D-2)\alpha e^{-\lambda\phi}(e^{\lambda\phi}-1)^{D-1}\times\cr
&&\qquad\Big(2(D-1)\big(1-\mu+(1+\mu)e^{\lambda\phi} \big)e^{-\fft{\phi}{\lambda(1+\mu)}}\cr
&&-e^{-\lambda\mu\phi}\big[(D-1)\big(1-\mu+(1+\mu)e^{\lambda\phi} \big)^2+(\mu^2-1)(e^{\lambda\phi}-1)^2\big]\times\cr
&&\qquad {}_2F_1\big[D-1,\ft 12 D(1-\mu),D,1-e^{\lambda\phi} \big]  \Big)\,,\nn
\eea
where $\lambda^2=2/((D-2)(1-\mu^2))$.  We find that analytical charged solution also exists provided that $Z(\phi)$ is given by
\be
Z^{-1}=\gamma e^{-\lambda\mu\phi}\Big(2(D-4)(1-\mu^2)+(1-\mu)\big(D-2-(D-4)\mu\big)
e^{-\lambda\phi}+(1+\mu)\big(D-2+(D-4)\mu\big)e^{\lambda\phi} \Big)\,.
\ee
The dilaton and the metric function $f$ are then given by
\bea
\lambda\phi &=& \log \big( 1 + \fft{q}{r}\big)\,,\cr
f&=&r^{1+\mu}(r+q)^{1-\mu}\Big(g^2-\fft{\alpha q^{D-1}}{r^{D-1}}{}_2F_1\big[D-1,\ft 12 D(1-\mu),n,-\ft qr\big] \Big)\cr
&&+\fft{2\gamma Q^2}{(D-2)\big(r^{1+\mu}(r+q)^{1-\mu} \big)^{D-3}}\,,
\eea
From the expression for the field strength
\be
F=-\fft{Q}{Z(\phi)r^{D-2}}\Big(1+\fft{q}{r}\Big)^{\ft 12(D-2)(\mu-1)} dr\wedge dt\,,
\ee
we obtain the gauge potential $A=a(r) dt$, where
\be
a=\fft{2\gamma Q}{r^{DD-2}} (2r + (1+\mu) q) \big(1 + \fft{q}{r}\big)^{\fft12((D-4)\mu +2-D)}\,.
\ee
The neutral solution with $\mu\ne0$ was obtained in \cite{Acena:2013jya} in a different coordinate system.  The neutral $\mu=0$ solution was given in \cite{Fan:2015tua} using the same coordinates. It is easy to calculate that the mass of the solution is given by
\be
M=\fft{\alpha (D-2) q^{D-1}}{16\pi}\,.
\ee
It is straightforward to verify that the first law of the thermodynamics $dM=TdS + \Phi_e dQ_e$ holds, where $T,S$ can be calculated with the standard technique and
\be
Q_e=\fft{Q}{16\pi} \Sigma\,,\qquad \Phi_e=a(r_0)\,,
\ee
with $r_0$ being the location of the horizon.  These thermodynamics quantities satisfy the generalized Smarr relation
\be
M=\fft{D-2}{D-1} (TS +\Phi_e Q_e)\,.\label{gensmarr}
\ee
Note that this is not the usual Smarr relation (\ref{usualsmarr}) which breaks down in all AdS black holes unless one treats the cosmological constant as a thermodynamical variable.  This generalized Smarr relation is a consequence of the scaling symmetry associated with planar AdS black holes \cite{Liu:2015tqa}.
It was shown in \cite{Liu:2015tqa} that this generalized Smarr formula is directly related to the the universal bound \cite{Policastro:2001yc,KSS} of the viscosity to the entropy ratio in the AdS/CFT correspondence.

\subsection{Class 2}

The class-2 solutions are specified by the metric ansatz
\be
ds^2=-\sigma^2 fdt^2+\frac{dr^2}{f}+r^2dx_i^2\,,
\ee
The neutral scalar  hairy black holes were constructed in \cite{Fan:2015tua}, with the scalar potential given by
\bea
V &=&-2g^2(D-2)\mu (\nu-\lambda^2\phi^2)e^{2\lambda^2\phi^2}\cr
&& -(D-1)(D-2)\alpha e^{\lambda^2\phi^2} \Big(\phi^{2\nu} + \lambda^{-2\nu} (\nu-\lambda^2\phi^2)
e^{\lambda^2\phi^2}\big(\Gamma(\nu,\lambda^2\phi^2)-\Gamma(\nu)\big)\Big)\,,
\eea
where $\lambda=\sqrt{\ft{\mu}{4(n-2)}}\,,\nu=\fft{n-1}{2\mu}$.  The charged planar black hole can also be constructed analytically provided that $Z$ is given by
\be
Z^{-1}=\gamma\big(\phi^2+\ft{2(n-3)(n-2)}{\mu^2}\big)e^{2\lambda^2\phi^2}\,,
\ee
The solution is then given by
\bea
\phi &=& \big(\fft{q}{r}\big)^\mu\,,\qquad F=\fft{Q e^{-\lambda^2 \phi^2}}{2Z r^{D-2}} dr \wedge dt\,,\qquad
\sigma = e^{-\lambda^2 \phi^2}\,,\cr
f&=& e^{2\lambda^2\phi^2} r^2 \Big(g^2 + \alpha\,\nu\,\lambda^{-2\nu} \big(\Gamma(\nu, \lambda^2 \phi^2)-\Gamma(\nu)\big)+\fft{\gamma Q^2}{\mu^2r^{2(D-2)}}\Big)\,.
\eea
It is straightforward to verify that the first law of thermodynamics and the generalized Smarr relation (\ref{gensmarr}) both hold.

\section{Conclusions}

In this paper, we considered a class of Einstein-Maxwell-Dilaton theories where the dilaton coupling $Z(\phi)$ to the Maxwell field is not the usual exponential function, as in a typical supergravity theory. There are three criteria for us to determine the function $Z$.  The first is that it should provide a toy model for string loop effects so that the extremal solution has the smooth AdS$\times$Sphere near-horizon geometry.  The second is that
the theory admits two different black holes for given mass and charges, providing a counter example of the no-hair theorem.  Both of the above criteria can be satisfied when $Z(\phi)$ has a stationary point. The third is that we can construct exact solutions for these black holes.

The theories we constructed indeed admit in general two types of black holes. One is the usual RN black hole, for which the dilaton scalar is a fixed constant. For the same mass and charge, there can exist also another charged black hole, in which the dilaton field varies.  We analysed the global properties and obtained the first law of black hole thermodynamics, and compared the differences between the new black hole with the RN solution.  We found that some scalar potential could be introduced and exact solutions of charged AdS black holes were obtained.  We generalized the black holes to black $p$-branes.

     We further obtained a class of new charged AdS planar black holes and
show that they all satisfy the generalized Smarr relation, which can be viewed as the bulk dual of the viscosity/entropy ratio of the boundary field theory.

\section*{Acknowledgement}

 Z.-Y.~Fan is supported in part by NSFC Grants NO.10975016, NO.11235003 and NCET-12-0054; the work of H.L.~is supported in part by NSFC grants NO.11175269, NO.11475024 and NO.11235003.

\end{document}